\documentclass[10pt,conference]{IEEEtran}

\usepackage{cite}
\usepackage{amsmath,amssymb,amsfonts}
\usepackage{algorithmic}
\usepackage{graphicx}
\usepackage{textcomp}
\usepackage{xcolor}
\usepackage{color, colortbl}
\usepackage{url}
\usepackage[T1]{fontenc}
\usepackage[utf8]{inputenc}
\usepackage{moreverb}
\usepackage{xspace}
\usepackage[
backgroundcolor=blue!10]{todonotes}
\usepackage{framed}
\usepackage[normalem]{ulem}
\usepackage{adjustbox}
\usepackage{array}
\usepackage{multirow}
\usepackage{hhline}
\usepackage{booktabs}
\usepackage{flushend}
\usepackage{subfig}
\usepackage{dblfloatfix}
\usepackage{listings}
\usepackage[framemethod=TikZ]{mdframed}

\definecolor{green}{rgb}{0,0.6,0}
\definecolor{codegray}{rgb}{0.5,0.5,0.5}
\definecolor{codepurple}{rgb}{0.58,0,0.82}
\definecolor{lightgray}{rgb}{.9,.9,.9}
\definecolor{darkgray}{rgb}{.4,.4,.4}
\definecolor{purple}{rgb}{0.65, 0.12, 0.82}
\definecolor{patchred}{RGB}{255, 191, 155}
\definecolor{patchgreen}{RGB}{164, 242, 164}
\definecolor{patchyellow}{RGB}{255, 237, 149}
\definecolor{royalazure}{rgb}{0.0, 0.22, 0.66}
\definecolor{blue2}{rgb}{0.54, 0.81, 0.94}
\definecolor{lightgray}{gray}{0.9}

\lstdefinelanguage{JavaScript}{
  keywords={typeof, new, true, false, catch, function, return, null, catch, switch, var, if, in, while, do, else, case, break},
  keywordstyle=\color{blue}\bfseries,
  ndkeywords={class, export, boolean, throw, implements, import, this},
  ndkeywordstyle=\color{darkgray}\bfseries,
  identifierstyle=\color{black},
  sensitive=false,
  comment=[l]{//},
  morecomment=[s]{/*}{*/},
  commentstyle=\color{green}\ttfamily,
  stringstyle=\color{red}\ttfamily,
  morestring=[b]',
  morestring=[b]"
}

\newcolumntype{^}{!{\vrule width 1pt}}
\newcommand{\srct}[1]{\textbf{\tiny\ttfamily#1\rmfamily}}
\newcommand{\sourcecode}[1]{\ttfamily#1\rmfamily}
\newcommand{\srcadd}[1]{\colorbox{green!40}{#1}}
\newcommand{\srcremove}[1]{\colorbox{red!30}{#1}}
\newcommand{\str}[1]{\textcolor{blue}{#1}}
\newcommand{\num}[1]{\textcolor{orange}{#1}}
\newcommand{\mycaption}[1]{\textbf{\vspace{-25pt}\begin{center}\small#1\normalsize\end{center}\vspace{5pt}}}

\def\BibTeX{{\rm B\kern-.05em{\sc i\kern-.025em b}\kern-.08em
    T\kern-.1667em\lower.7ex\hbox{E}\kern-.125emX}}

\begin{document}

\title{Exploring Plausible Patches Using Source Code Embeddings in JavaScript}

\author{
    \IEEEauthorblockN{
			Viktor Csuvik,
			Dániel Horváth,
			Márk Lajkó,
			László Vidács
		}
		\\\IEEEauthorblockA{
			\textit{MTA-SZTE Research Group on Artificial Intelligence}
			\\ University of Szeged, Szeged, Hungary
		}
		\\\{csuvikv,hoda,mlajko,lac\}@inf.u-szeged.hu
}

\maketitle

\begin{abstract}
Despite the immense popularity of the Automated Program Repair (APR) field, the question of patch validation is still open. Most of the present-day approaches follow the so-called Generate-and-Validate approach, where first a candidate solution is being generated and after validated against an oracle. The latter, however, might not give a reliable result, because of the imperfections in such oracles; one of which is usually the test suite. Although (re-) running the test suite is right under one's nose, in real life applications the problem of over- and underfitting often occurs, resulting in inadequate patches. Efforts that have been made to tackle with this problem include patch filtering, test suite expansion, careful patch producing and many more. Most approaches to date use post-filtering relying either on test execution traces or make use of some similarity concept measured on the generated patches. Our goal is to investigate the nature of these similarity-based approaches. To do so, we trained a Doc2Vec model on an open-source JavaScript project and generated 465 patches for 10 bugs in it. These plausible patches alongside with the developer fix are then ranked based on their similarity to the original program. We analyzed these similarity lists and found that plain document embeddings may lead to misclassification - it fails to capture nuanced code semantics. Nevertheless, in some cases it also provided useful information, thus helping to better understand the area of Automated Program Repair.
\end{abstract}


\section{Introduction}
\label{sec:introduction}

Automated Program Repair (APR) in the research direction has recently reached new heights with promising results recorded \cite{Monperrus2018bib,LeGoues2019apr}. Nothing shows this better than the many program repair tools that have been implemented and the papers that has been published~\cite{Martinez2014,Tufano2018,Mahajan2018a,Martinez2017,Gupta2016,Xuan2017}. Many of these tools follow the Generate-and-Validate (G\&V) approach, which first localizes the exact location of the bug in the source code, typically with the help of the test suite, creating a list of the most \emph{suspicious} parts of the program to repair. The intuition is that if the suspicious parts are repaired, the program will work correctly. After localization patch candidates are generated usually by search-based methods. To validate the generated patches, an oracle is needed which can reliably determine if the repair is really correct and can detect if over- or underfitting has occurred. Test suites are widely used as affordable approximations of this oracle in the APR field, therefore (re-)running the tests validates whether the repair process was successful or not. A program is marked as a plausible patch, if it passes all the available test cases. This latter condition gives no assurance that the program is \emph{correct}, since over- and underfitting~\cite{Le2018,Qi2015,Smith2015} often occurs, resulting in inadequate patches.

Recent works~\cite{Tian2020,Csuvik2020::IBF,Le2019,Tufano2019,Wang2019,Xiong2018} on patch correctness assessment show the importance of it. Most of these works utilize a similarity concept (i.e. Doc2Vec~\cite{Le2014}, Bert~\cite{Devlin2019}, code2vec~\cite{Alon2018}) and compute the similarity of software artifacts (e.g. test case execution traces or generated patches). Deciding the correctness of a candidate patch~\cite{Assiri2017,Gazzola2018} is one of the current challenges in automated program repair, and today considered as an open question~\cite{Gazzola2019}. However some repair tools try to generate only real correct patches using carefully-designed (e.g., fine-grained fix ingredients) repair operators or test-suite augmentation, but none of these approaches are able to eliminate overfitting completely~\cite{Yang2017,Wen2018}.

\begin{figure}[t]
	\centering
		\includegraphics[height=0.9\columnwidth]{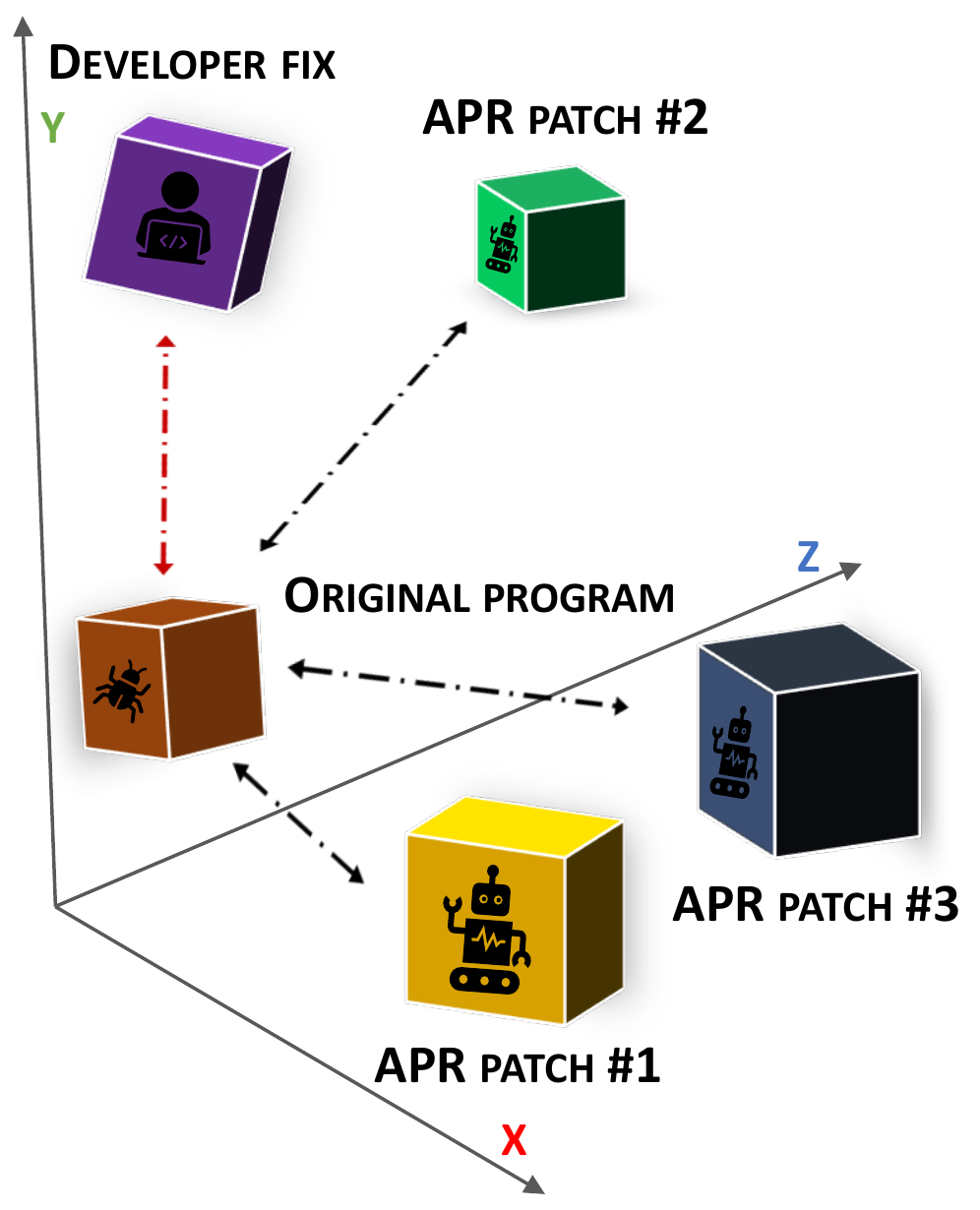}
	\caption{A high level overview of the proposed approach}
	\label{fig:overview}
\end{figure}

In this work instead of providing another technique for testing the correctness of patches, we are looking at whether similarity is an appropriate approach to tackle with the patch correctness problem. To do so, we examine not just the generated plausible patches, but the developer fix as well. In Figure~\ref{fig:overview} one can observe a similarity-based approach, where plausible patches are compared to the original program. Using source code embeddings allow us to place these textual documents into some high-dimensional space, where usual similarity measures can be applied. For example in Figure~\ref{fig:overview} one can observe, that APR patch \#2 lies closer to the developer patch than the other patches, thus it might be better in some way. Of course in higher dimensions, the similarity metric should be well defined, and the exact meaning of each dimension can not be interpreted in most cases. In this work we measured similarity between the generated patches and the original program and also included the developer fix in the comparison process, thus seeking for answers to the following research questions:

\noindent\textit{\textbf{RQ1:} How effectively can the similarity-based approach be used to validate plausible patches?}

\noindent\textit{\textbf{RQ2:} Do correct patches always show great similarity with the original program?}

Our main motivation is to create an exploratory analysis about the similarity-based patch validation. The assumption of similarity-based techniques is that the \emph{correct} program is more similar to the original one than other candidates. It comes from the perception that the current techniques mostly create single line code modifications, thus leaving most of the original source code intact. When repairing a program, it is preferable to construct patches which are simple and readable that is, to be more similar to the original code base~\cite{Mechtaev2015}. A well experienced software developer would share the same goal: fix the issue in a way and style that it integrates with the original code base. This is because responsible software maintainers would review and inspect a patch carefully before accepting it~\cite{Scacchi2006} – which occurs only if they judge that the patch is correct and safe~\cite{Mechtaev2015}. Although textual and structural similarity does not imply that the constructed patch is \emph{simple} for human software developers, Csuvik et al.~\cite{Csuvik2020::IBF} found that in some way, similarity indicates understandability and if a patch is more understandable, its chance of being \emph{correct} is higher. Their approach used the same ranking method as ours, except our experiments are conducted on JavaScript. The contributions of the paper include (1) quantitative and qualitative analysis of plausible patches for JavaScript (2) adaptation of a similarity-based patch filtering method (3) manual annotation of 465 patches. Although we do not present a new technique that can be used for patch filtering, we are adapting one to JavaScript and analyzing its usability in depth.

In this paper we used bugs from the BugsJS dataset~\cite{Gyimesi:ICST:2019:BugsJS} which containes 453 reproducible JavaScript bugs from 10 open-source Github projects. The tool we used is an adaptation of the original GenProg~\cite{Weimer-ICSE-2009-Genprog} method supplemented with JavaScript-specific repair operators\footnote{The complete description of the tool is beyond the scope of the current paper, the interested reader may find more information about it in https://github.com/GenProgJS.}. We used the generated plausible patches for 10 distinct bugs all from the Eslint project. We restricted our experiments to the Eslint project because it is the largest project in the BugsJS dataset, it contains the most single-line errors.

The paper is organized as follows. In the next section we present the experiment setup including the applied method (Section~\ref{sec:method}), the used embedding technique (Section~\ref{sec:doc2vec}) and the sample plausible patches (Section~\ref{sec:bugsjs}). Evaluation and analysis are presented in Section~\ref{sec:results}. Related work is discussed in Section~\ref{sec:related}, and we conclude the paper in the last section.

\section{Experimental setup}
\label{sec:experiment_setup}

\begin{figure}[!b]
	\centering
		\includegraphics[width=0.8\columnwidth]{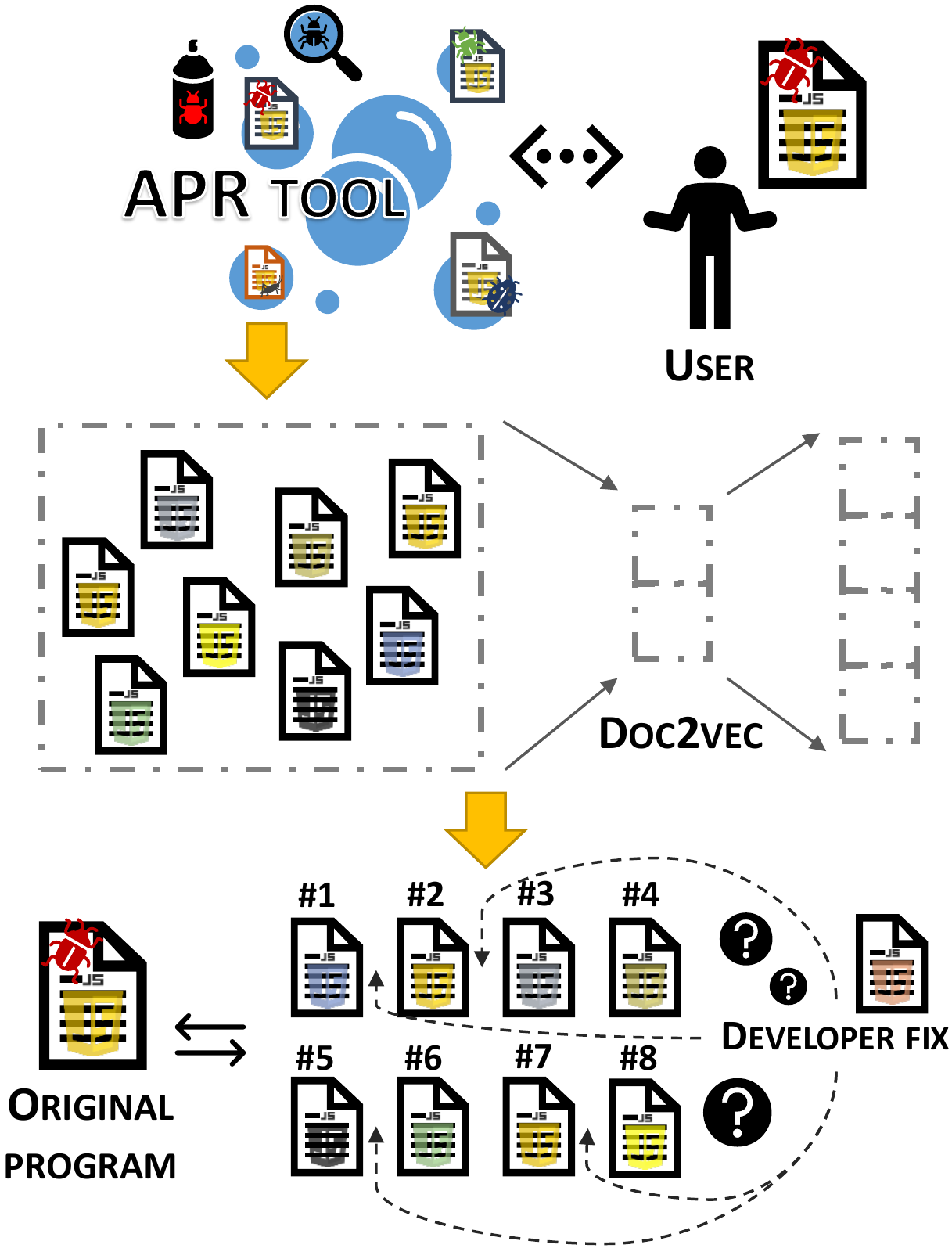}
	\caption{Illustration of the implemented process}
	\label{fig:process}
\end{figure}

In this section we describe the used approach to determine the usefulness of similarity based patch validation. A high-level overview of the proposed process can be found in Figure~\ref{fig:process}. First an APR tool creates plausible patches, usually more than one. In our case the tool always ran for 30 generations resulting in a high number of plausible patches for each bug. From the original program and from the generated potentially fixed programs, we extract the faulty line and a small environment of it - this snippet of code will serve as a basis for calculating similarity. After a Doc2Vec model is trained, for every code snippet an N dimensional vector is created on which one can measure similarity. These vectors contain information about the meaning, environment, and context of a word or document. The generated plausible patches then lined up alongside with the developer fix, based on the similarities calculated previously. Based on this list we can analyze which version of the fixed program is the most similar to the original one - the one created by an APR tool or a developer fix.

In the next subsections the evaluation procedure is presented first, then the method used to measure similarity, and finally the used plausible patches are described.

\subsection{Method}
\label{sec:method}

For the investigated bugs, determining, whether a fix for it is correct or not, was usually quite easy. In general to answer this question again can be a challenging, but in our case the generated patches were one-liners and usually it was clear whether the APR tool generated an overfitted patch or a correct one. Although determining which fix is better than the other one can be tricky. A rule-of-thumb can be that the simpler a patch the better. This, of course, supports similarity-based validation, but it may not be the right solution in all cases. Ranking plausible patches alongside with the developer patch may also point out that a human written patch differs greatly from the original source code. This does not necessarily mean that the developer made a mistake: it might be that he adapts a new approach that was never used before in the code base, or simply made a refactoring of considerable size.

Our main question is, whether it is true — from the perspective of Doc2Vec — that the developer fix lies close to the original program. Current state-of-the-art APR applications still fail to repair real complex issues, thus the demand for simple patches may be desirable. To measure the quality of the ranking, we used the Normalized Discounted Cumulative Gain (\emph{nDCG}) metric~\cite{ndcg}, which is computed as:

\begin{equation}
	\label{eq:NDCG}
		nDCG_p=\frac{DCG_p}{IDCG_p}
\end{equation}

Where \emph{DCG} stands for Discounted Cumulative Gain, \emph{IDCG} stands for Ideal \emph{DCG} and \emph{p} is a particular rank position. \emph{DCG} measures the usefulness, or gain, of a document based on its position in the result list. \emph{IDCG} basically is the maximum possible \emph{DCG} value that can be achieved on a ranked list - this is done by sorting all relevant documents in the corpus by their relative relevance. Since the similarity lists vary in length (the number of plausible patches is different for each bug), consistently comparing their performance with \emph{DCG} is not possible. To be able to compare these lists, cumulative gain at each position for a chosen value of \emph{p} should be normalized, thus resulting in the \emph{nDCG} metric defined above in Equation~\ref{eq:NDCG}. The definition of \emph{DCG} and \emph{IDCG} is presented in Equation~\ref{eq:DCG} and Equation~\ref{eq:IDCG} respectively.

\begin{tabular}[c]{cc}
	\parbox{.4\columnwidth}{\centering
		\begin{equation}
			\label{eq:DCG}
				DCG_p=\sum_{i=1}^{p} \frac{2^{rel_i} - 1}{log_2 (i + 1)}
		\end{equation}}
& \parbox{.4\columnwidth}{ \centering
		\begin{equation}
			\label{eq:IDCG}
				IDCG_p=\sum_{i=1}^{|REL_p|}\frac{2^{rel_i} - 1}{log_2 (i + 1)}
		\end{equation}} \\
\end{tabular}

The $rel_i$ is the graded relevance of the result at position $i$. Since the similarity values give us the ordering, each item in the list should have another value which validates its placement. We manually checked each and every generated bug and categorized them based on their relevance. The following relevance scores were introduced:
\begin{itemize}
	\item 3: the developer fix always has the highest relevance, in ranking the most favorable is when this patch comes first
	\item 2: the patch syntactically matches the developer fix - we use the term syntactic match when the codes are the same character by character, apart from white spaces
	\item 1: it is semantically identical to the developer fix - that is, the two source codes have the same semantical meaning, but there may be character differences
	\item 0: we were uncertain about the patch
	\item -1: the patch is clearly incorrect (e.g. syntactic errors)
\end{itemize}
In addition to these, intermediate categories are also conceivable: e.g. -0.5 would mean that a patch is probably incorrect, but we were not sure about that, because the lack of domain knowledge about the system examined. Two experienced software developers separately annotated the generated patches, they did not have the chance to influence each other. In cases where individual scores differed a third expert decided on the correctness of the patch. These annotated relevance scores are available in the online appendix of this paper\footnote{https://github.com/RGAI-USZ/JS-patch-exploration-APR2021}.

\subsection{Learning Document Embeddings}
\label{sec:doc2vec}
For every bug a small environment of the faulty line was selected, and this was embedded using Doc2Vec. This code environment includes the faulty line itself and three lines in front of and behind it. For our training data we tokenized this small code fragment with a simple regular expression, which separated words and punctuations, except for words with the dot '.' (member) operator. For a simple code example like: \textit{function foo () \{ return this.bar; \}} the tokenized version would be: \textit{ 'function', 'foo', '(', ')', '\{', 'return', 'this.bar', ';', '\}' }. Doc2Vec is a fully connected neural network that uses a single hidden layer to learn document embeddings (N dimensional vectors). We feed the input documents to this neural network, and it computes the embedding vectors, on which conceptual similarity can be measured. For our Doc2Vec configuration we used a vector size of 256, which basically is the dimensionality of the feature vectors. Window size of 5 was used, which tells Doc2Vec the maximum distance between the current and predicted word. Every word with a frequency of less than 2 was ignored. As for the training, the model was trained for 50 epochs. Every other parameter was left as default.
On the obtained embeddings (vectors containing real numbers) similarity is measured with the \emph{COS3MUL} metric, proposed in~\cite{Levy2014}. According to the authors positive words still contribute positively towards the similarity, negative words negatively, but with less susceptibility to one large distance dominating the calculation.

\subsection{Sample plausible patches}
\label{sec:bugsjs}

The BugsJS dataset~\cite{Gyimesi:ICST:2019:BugsJS} containes 453 reproducible JavaScript bugs from 10 open-source Github projects. The dataset contains multi-line bugs as well, which are beyond the scope of the current research. There are 130 single-line bugs, but not every one of them are "repairable", because of the lack of failed test cases. Thus, the number of single-line bugs, for which there is at least 1 failed test is 126 (and 94 only comes from the Eslint project). BugsJS features a rich interface for accessing the faulty and fixed versions of the programs and executing the corresponding test cases. These features proved to be rather useful for the comparison of automatically generated patches with the ones which were created by developers. We limited our experiments strictly to the Eslint project because it is the largest project in the BugsJS dataset, it contains the most single-line errors. The automatic repair tool which we used was able to repair 10 bugs from 94 in the Eslint project. Since the tool was configured to run for 30 generations in every case (so it does not stop at first when a fix is found), there was a high number of repair candidates in most cases of the runs as can be seen in Table~\ref{tab:bugsjs_repairs}. In the first column one can find the id of each bug and next to it how many plausible patches were generated to it. The two remaining columns show the original source code and a fix for it created by a developer.

\begin{figure}[b]
	\centering
		\includegraphics[width=1\columnwidth]{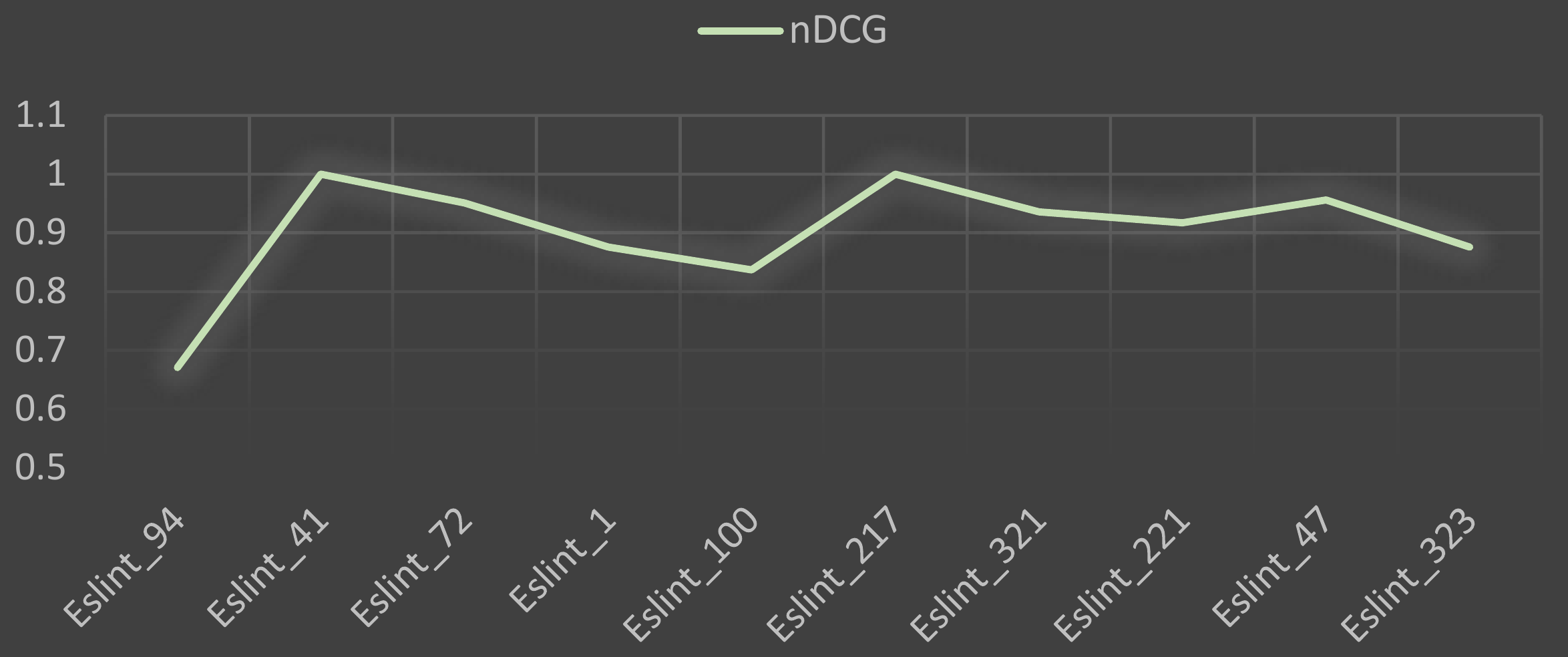}
	\caption{The nDCG metric value based on each of the examined bugs' ranked list}
	\label{fig:metrics}
\end{figure}

\begin{table*}[t]
	\centering
	\small
		\caption{Plausible patches and their corresponding developer fix in the Eslint project}
		\begin{tabular}{l^c|l|l}
		Bug Id & \#candidate & Original line & Developer fix \\
		\midrule
		\rowcolor{lightgray} Eslint 1 & 4 & \srct{if (name === \str{"Math"} || name === \str{"JSON"})} & \srct{if (name === \str{"Math"} || name === \str{"JSON"} \srcadd{|| name === \str{"Reflect"}})} \\
		Eslint 41 & 3 & \srct{end.column === line.length\srcremove{)}} & \srct{\srcadd{(}end.\srcadd{line === lineNumber \&\& end.}column === line.length\srcadd{));}} \\
		\rowcolor{lightgray} Eslint 47 & 3 & \srct{column: \srcremove{\num{1}}} & \srct{column: \srcadd{\num{0}}} \\
		Eslint 72 & 7 & \srct{loc: \srcremove{lastItem}.loc.end,} & \srct{loc: \srcadd{penultimateToken}.loc.end,} \\
		\rowcolor{lightgray} Eslint 94 & 14 & \srct{op.type === \str{"Punctuator"} \&\&} & \srct{\srcadd{(}op.type === \str{"Punctuator"} \srcadd{|| op.type === \str{"Keyword"})} \&\&} \\
		Eslint 100 & 12 & \srct{penultimateType === \str{"ObjectExpression"}} & \srct{\srcadd{(}penultimateType === \str{"ObjectExpression"} \srcadd{|| penultimateType === \str{"ObjectPattern"})}} \\
		\rowcolor{lightgray} Eslint 217 & 4 & \srct{if (!options || typeof \srcremove{option} === \str{"string"})} & \srct{if (!options || typeof \srcadd{options} === \str{"string"})} \\
		Eslint 221 & 221 & \srct{return \srcremove{parent.static};} & \srct{return \srcadd{false};} \\
		\rowcolor{lightgray} Eslint 321 & 5 & \srct{...loc.end.line !== node.loc.\srcremove{end}.line \&\&...} & \srct{...loc.end.line !== node.loc.\srcadd{start}.line \&\&...} \\
		Eslint 323 & 192 & \srct{else if (definition.type === \str{"Parameter"}} & \srct{else if (definition.type === \str{"Parameter"})} \\
		& & \srct{\srcremove{\&\& node.type === \str{"FunctionDeclaration"})}} & \\
		\midrule
		\textbf{Total} & 465 \\
		\end{tabular}
	\label{tab:bugsjs_repairs}
\end{table*}

\begin{figure*}[!t]
    \centering
    \subfloat[\centering Eslint 1]{{\includegraphics[width=0.40\columnwidth]{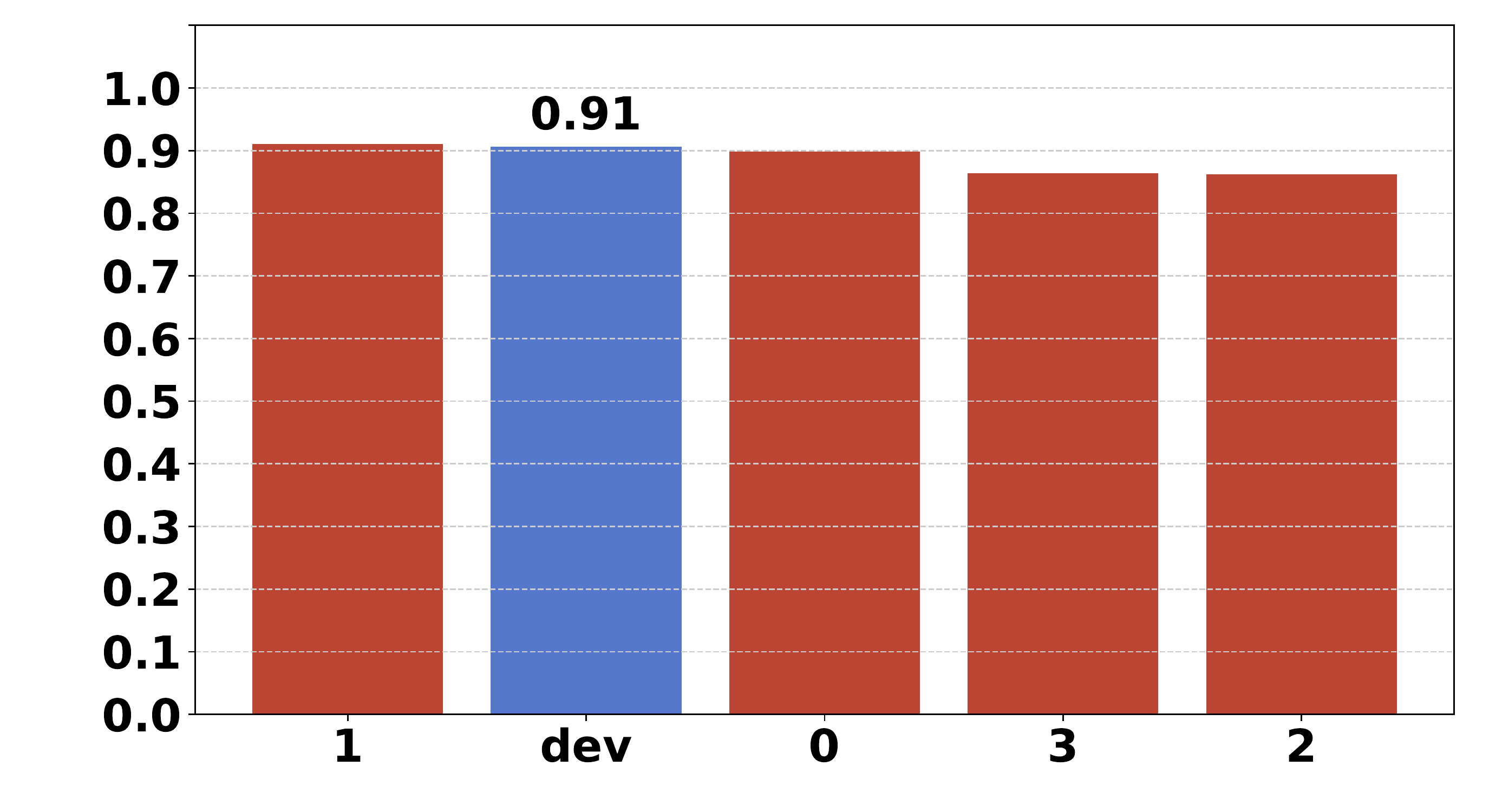} }}
    \subfloat[\centering Eslint 41]{{\includegraphics[width=0.40\columnwidth]{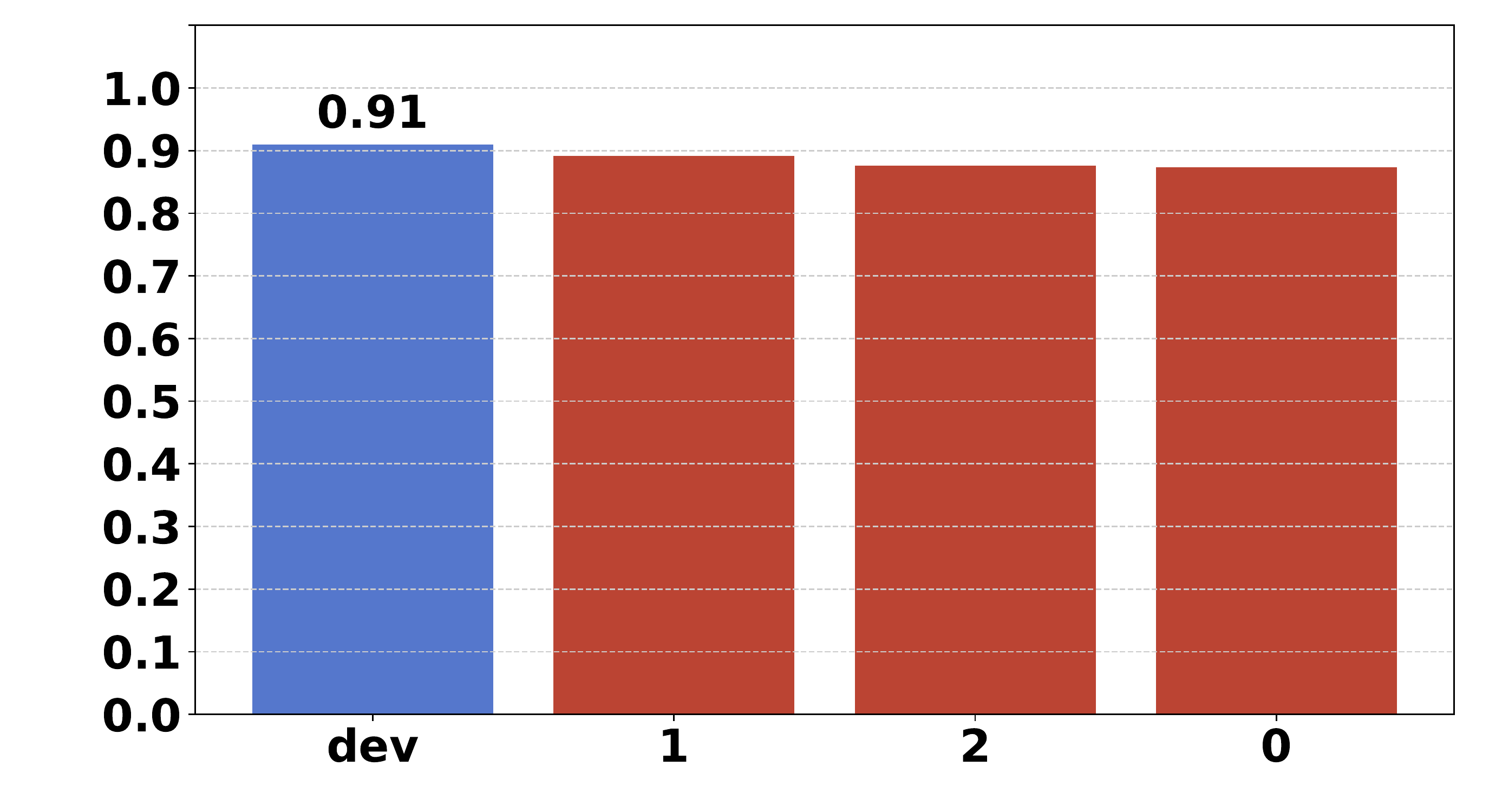} }}
    \subfloat[\centering Eslint 47]{{\includegraphics[width=0.40\columnwidth]{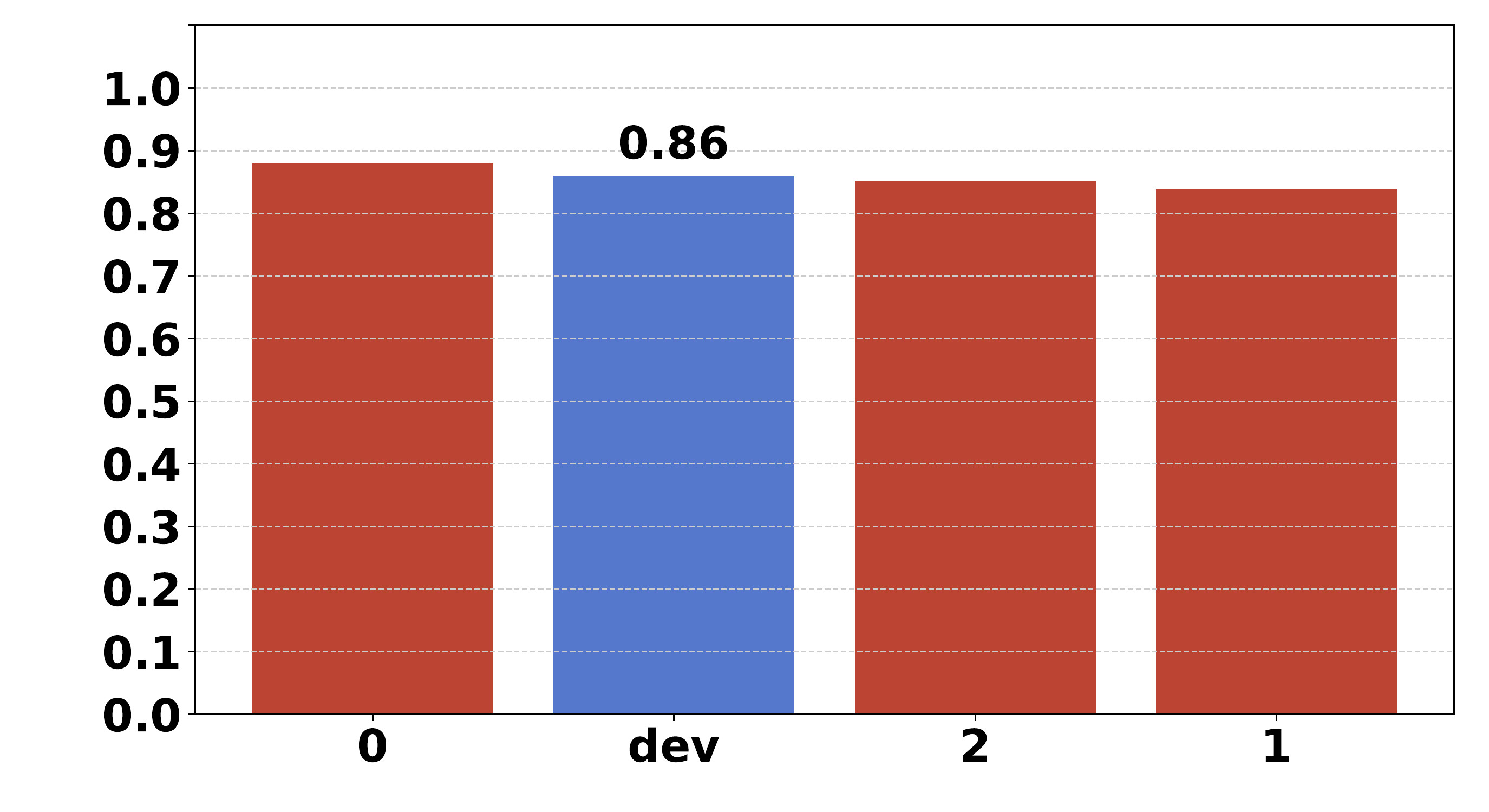} }}
    \subfloat[\centering Eslint 72]{{\includegraphics[width=0.40\columnwidth]{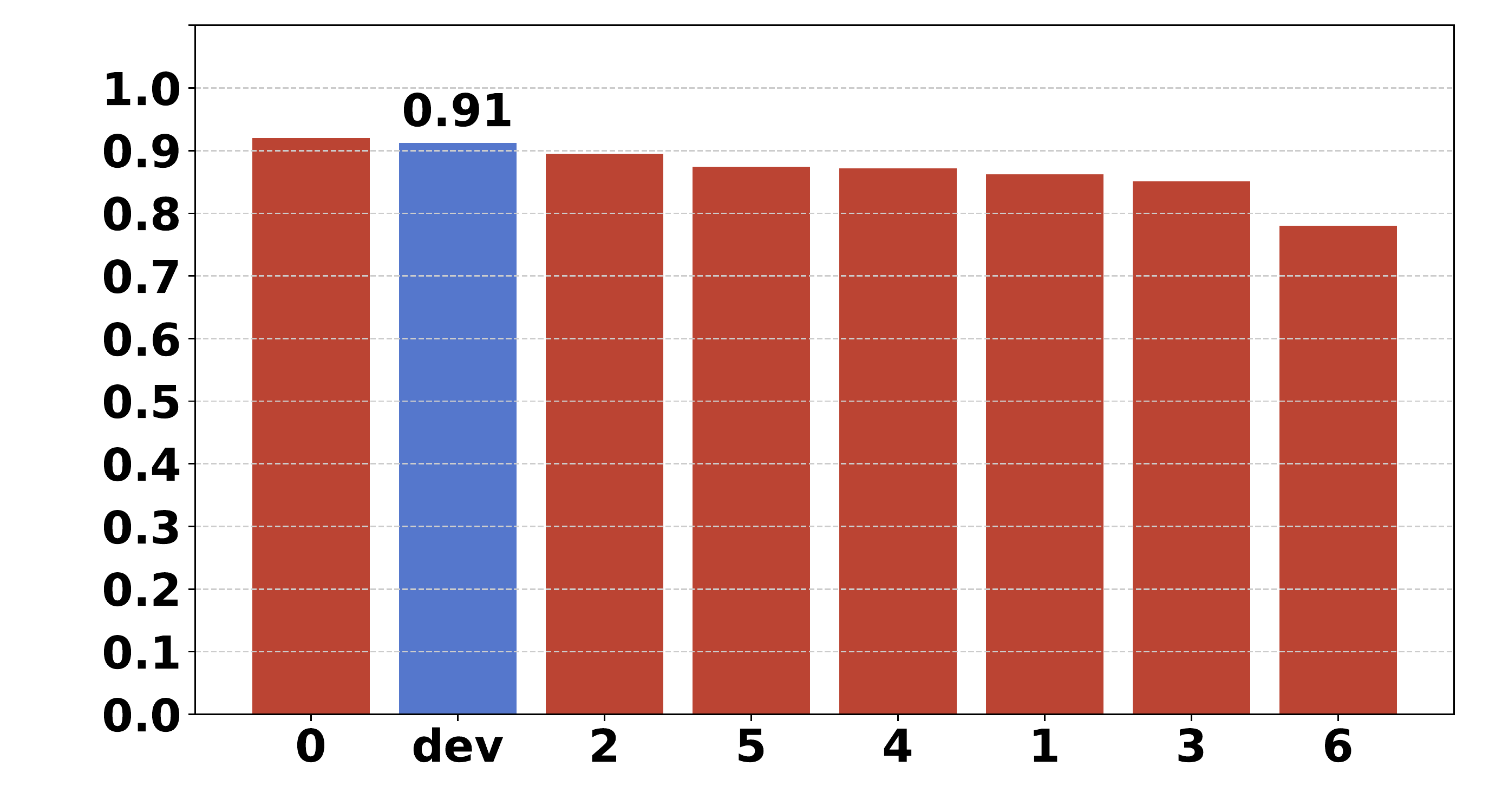} }}
    \subfloat[\centering Eslint 94]{{\includegraphics[width=0.40\columnwidth]{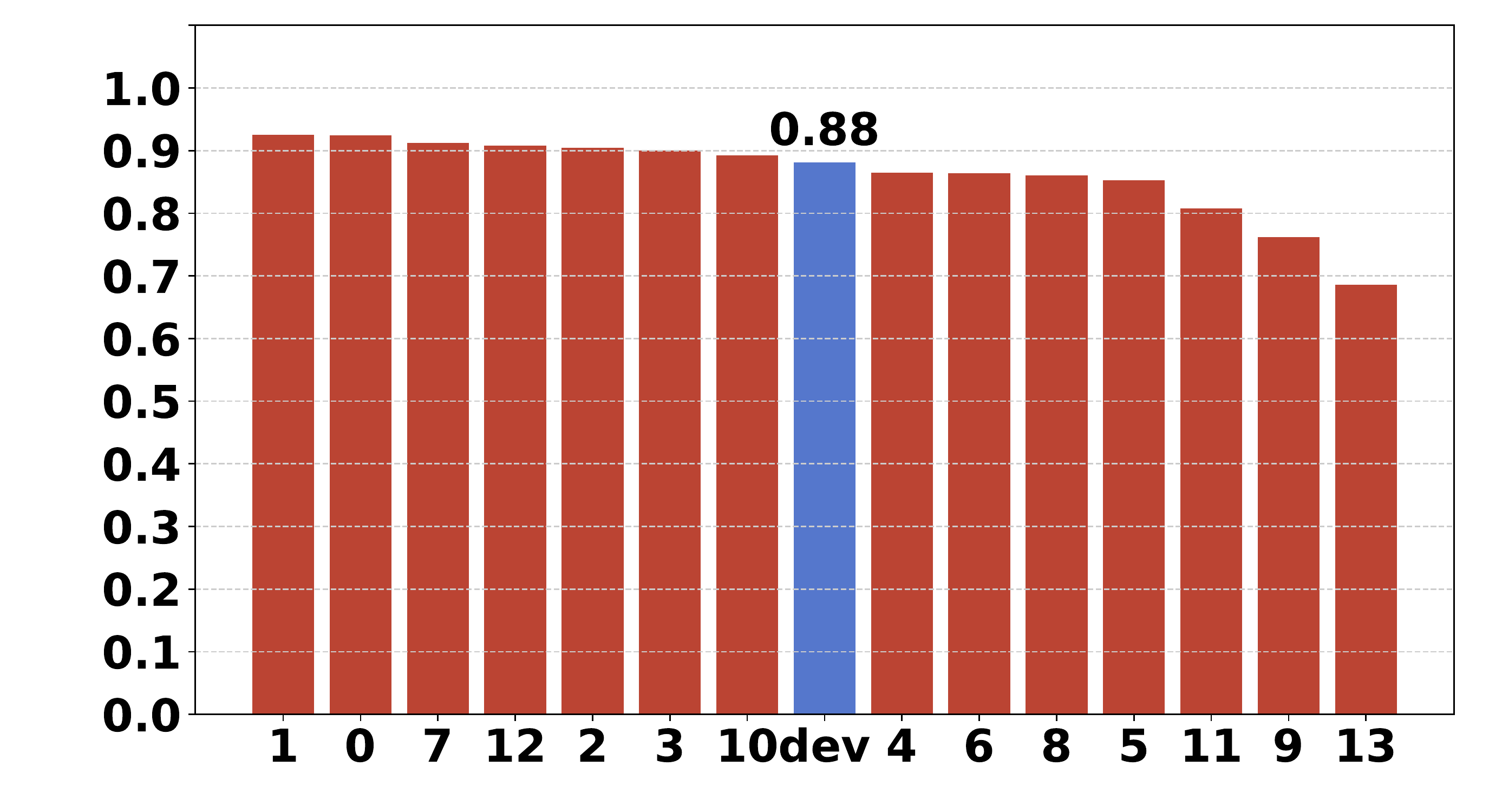} }}
		\qquad
		\subfloat[\centering Eslint 100]{{\includegraphics[width=0.40\columnwidth]{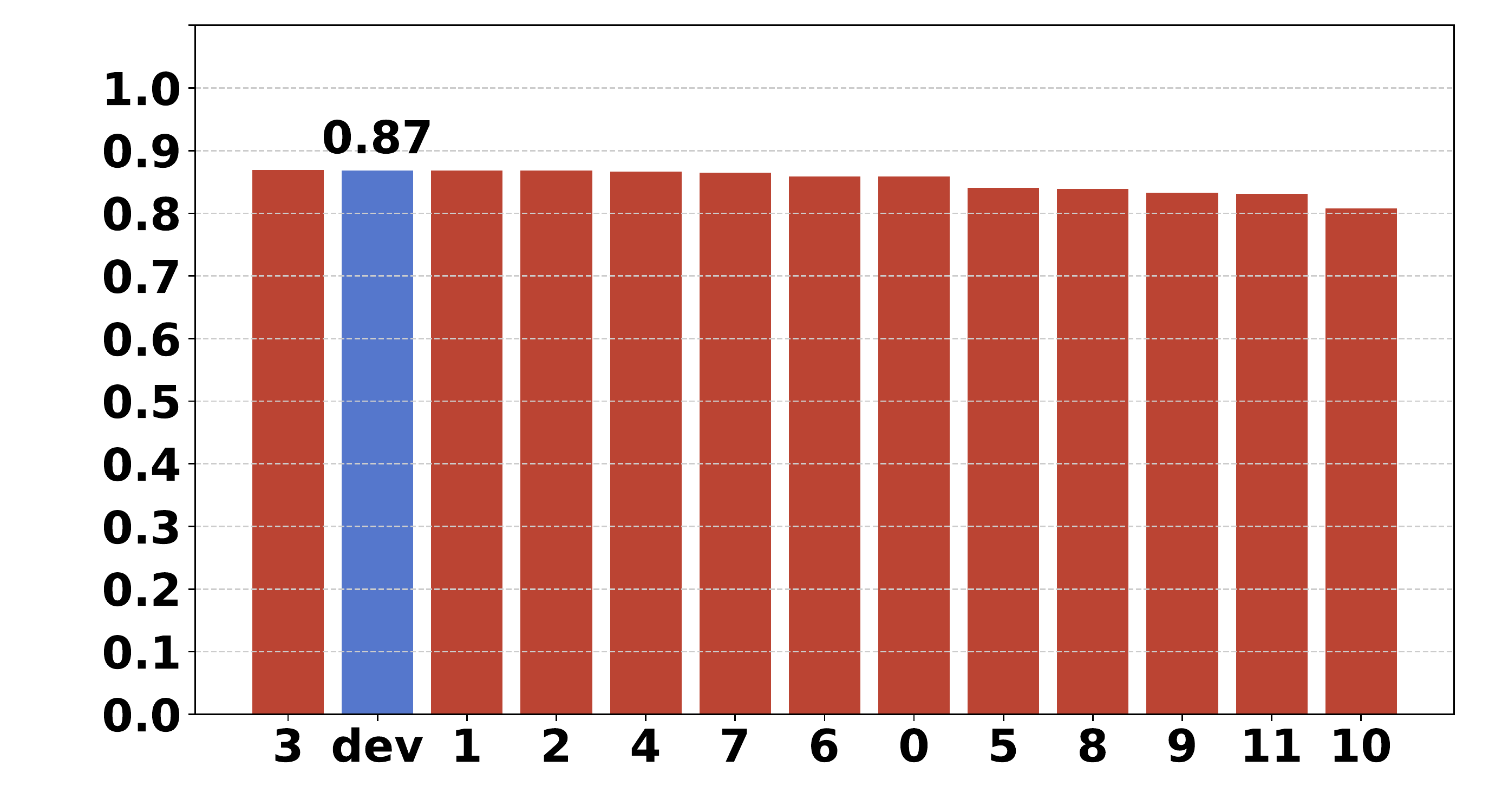} }}
    \subfloat[\centering Eslint 217]{{\includegraphics[width=0.40\columnwidth]{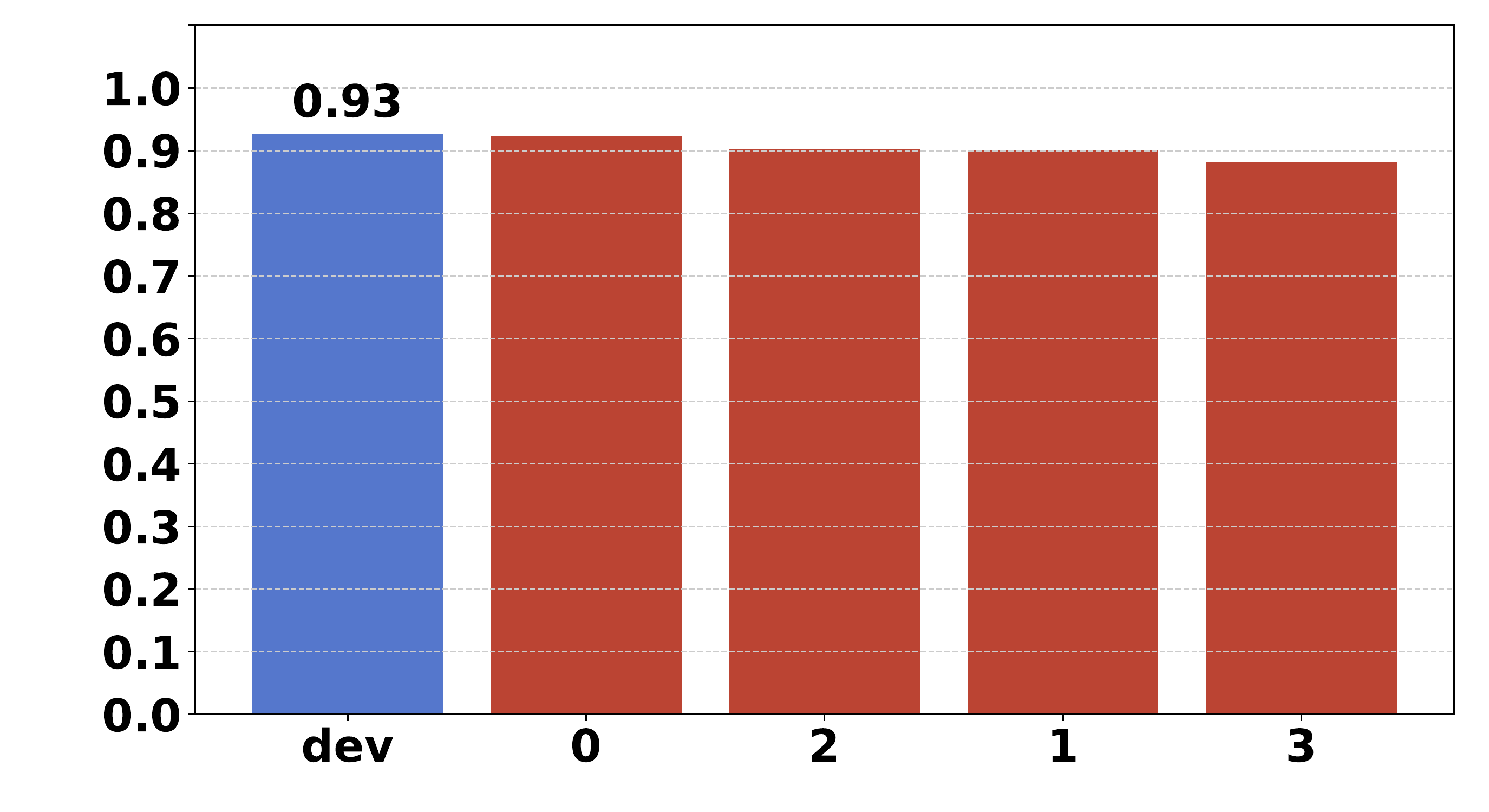} }}
    \subfloat[\centering Eslint 221]{{\includegraphics[width=0.40\columnwidth]{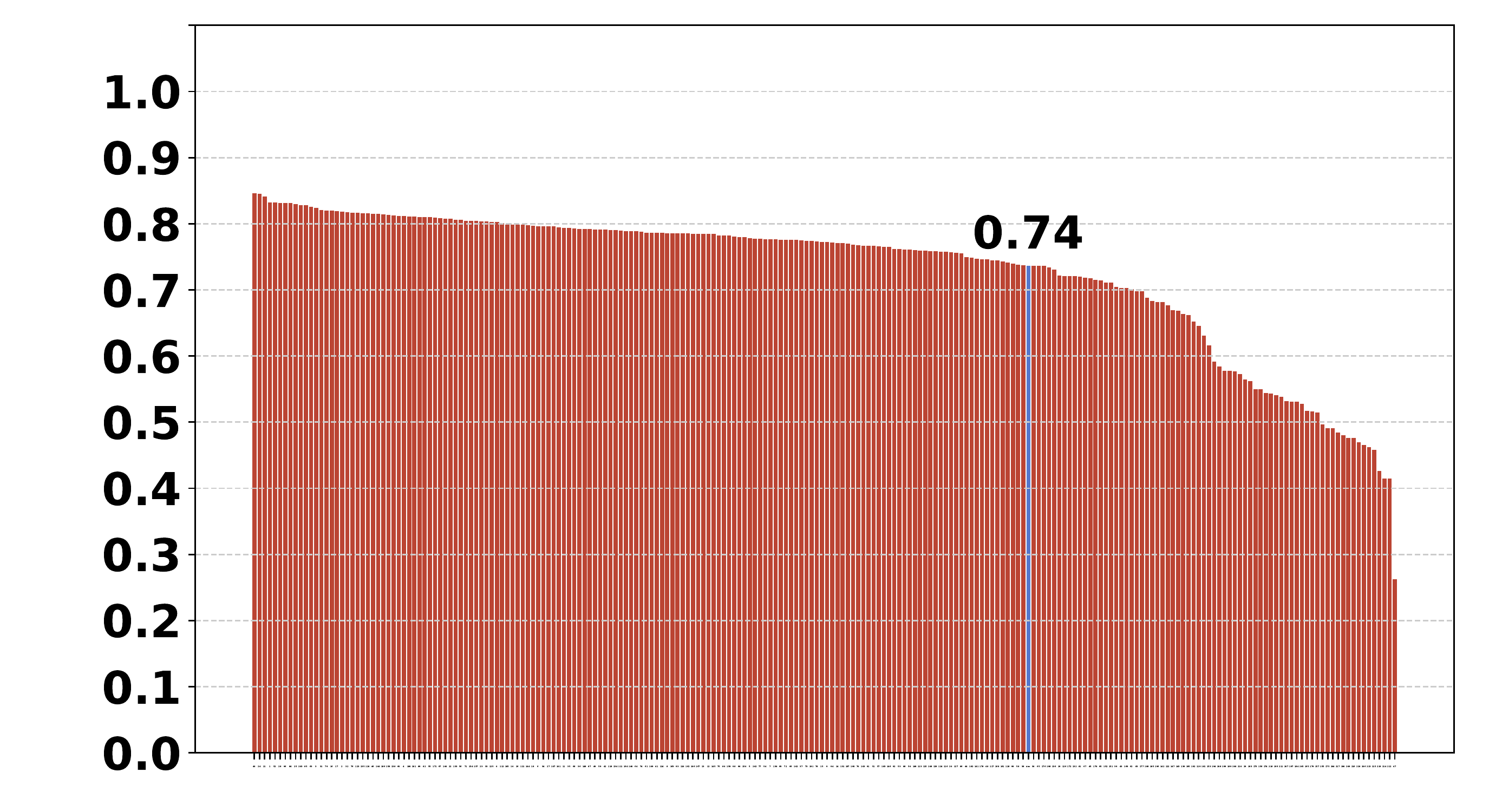} }}
    \subfloat[\centering Eslint 321]{{\includegraphics[width=0.40\columnwidth]{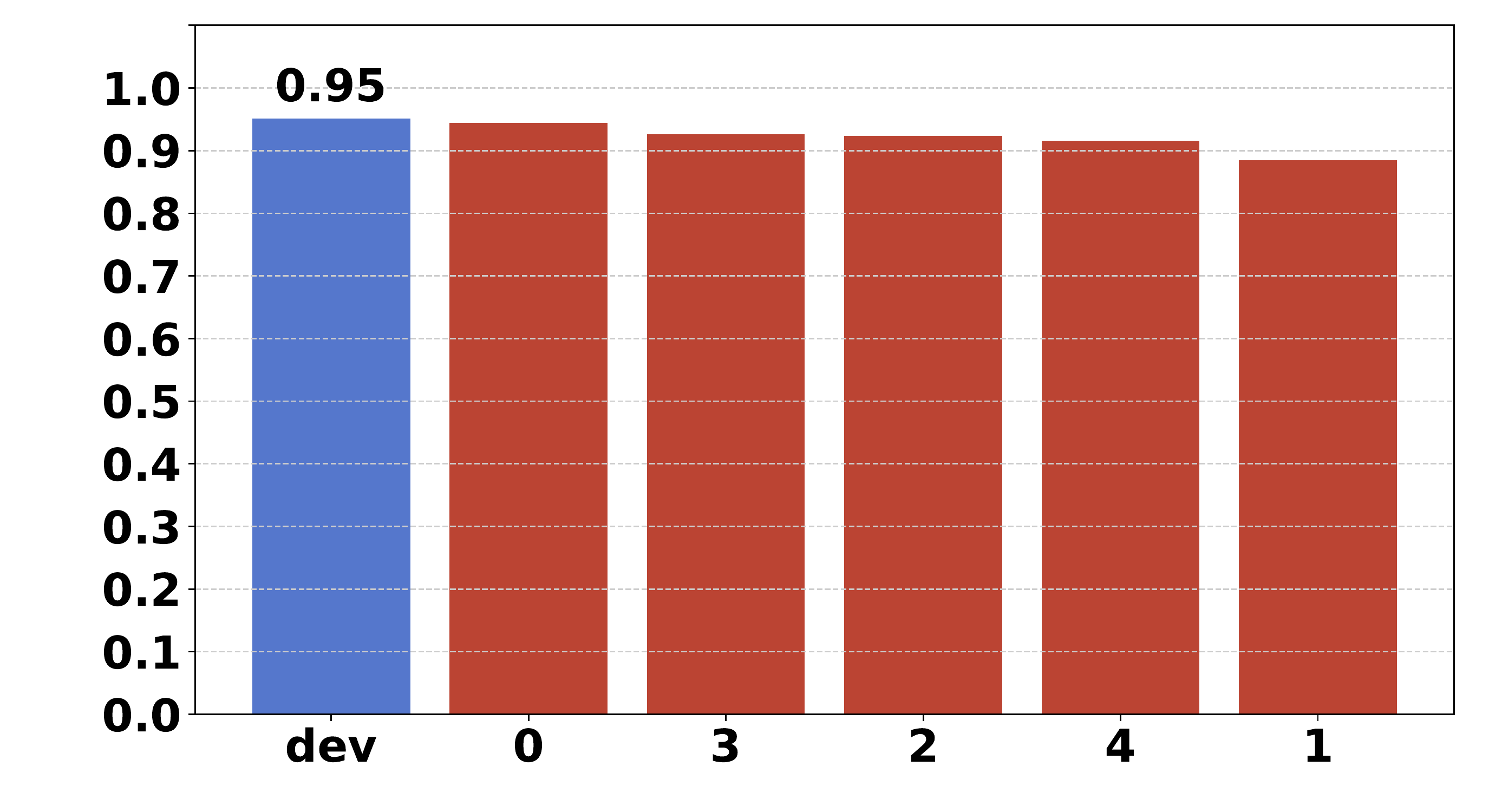} }}
    \subfloat[\centering Eslint 323]{{\includegraphics[width=0.40\columnwidth]{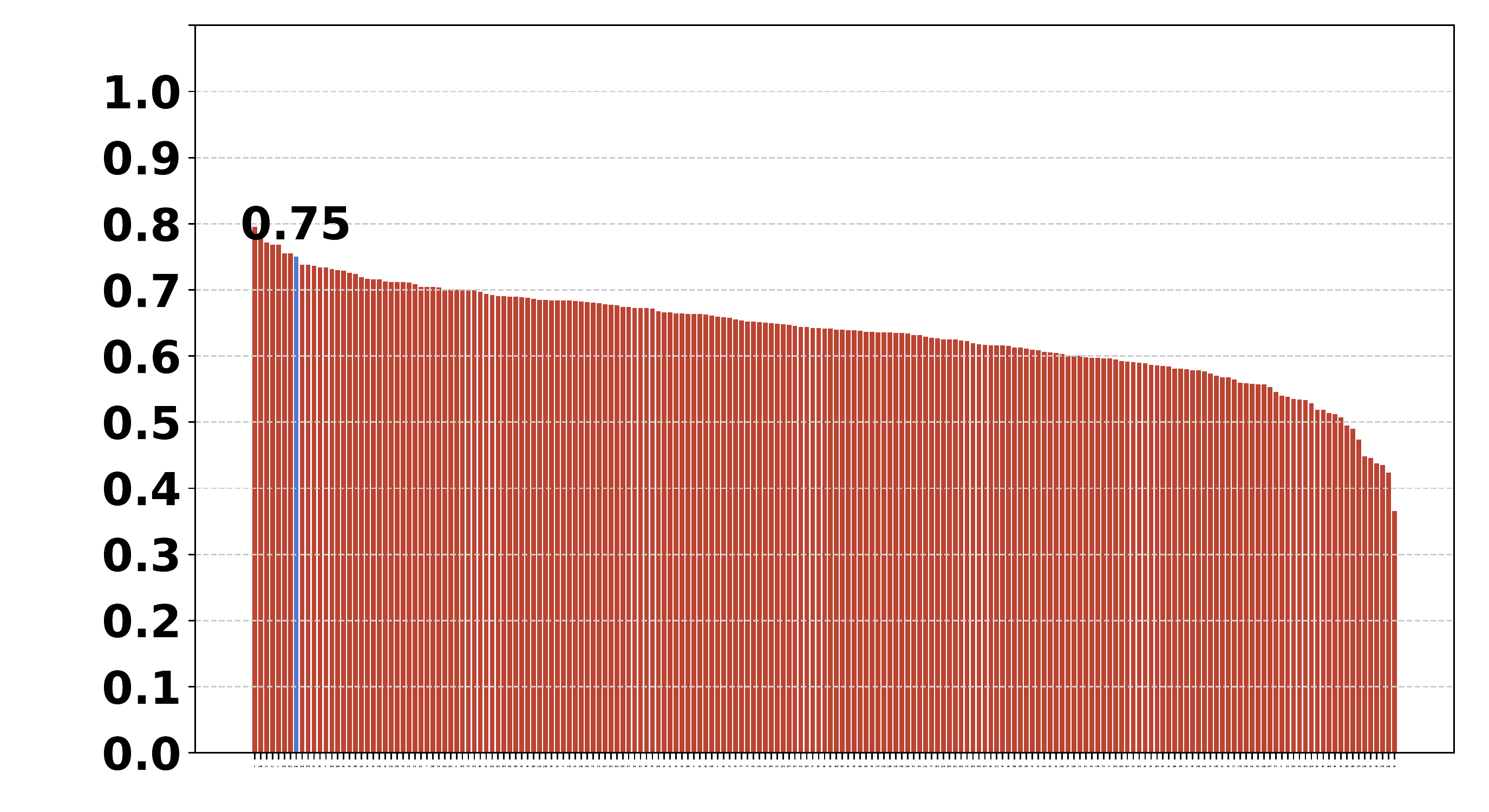} }}
		\qquad
		\centering{{\includegraphics[width=0.8\columnwidth]{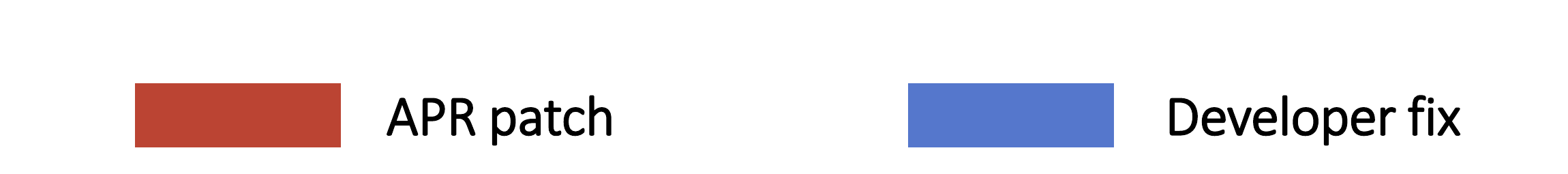} }}
    \caption{The developer fix and patches ranked based on their similarity to the original program}
    \label{fig:sims}
\end{figure*}

Each of the examined bugs and their corresponding fixes are one-liners, meaning that the modification that a developer (or APR tool) has to make to fix the bug only affects one line. We can see that there are bugs of different difficulty in Table~\ref{tab:bugsjs_repairs}: from quite simple where a number had to be replaced (Eslint 47), to quite complex where a conditional expression needed to be supplemented (Eslint 100). The number of generated candidates also varies greatly, this is due to the difficulty of the fixes and the random factor in the GenProg algorithm. In total 465 plausible patches were generated. We checked these patches and found that only three of these are syntactically identical to the developer fix (Eslint 47, Eslint 323 and Eslint 72), although many of them are semantically identical.

It is apparent that for Eslint 221 and Eslint 323 the number of plausible patches is orders of magnitude more than for any other. To explain it let us examine the nature of these bugs. The case of Eslint 221 is quite easy to understand: the return value should be \sourcecode{false}, making it rather simple to generate. We examined the generated patches and found that essentially anything would satisfy this criteria: in JavaScript \sourcecode{0}, \sourcecode{-0}, \sourcecode{null}, \sourcecode{false}, \sourcecode{NaN}, \sourcecode{undefined}, or the empty string (\sourcecode{""}) create an object with an initial value of \sourcecode{false}. On the other hand in case of Eslint 323 the high number of plausible patches is most probably because of the weak test suite. As we can see from Table~\ref{tab:bugsjs_repairs} the fix is not quite obvious, but after carefully inspecting the generated fixes we came to the conclusion that every modification on which the \sourcecode{if} condition evaluated to \sourcecode{true} successfully passed testing.

\section{Results and Discussion}
\label{sec:results}

In this section we evaluate and analyze the generated patches and similarity lists. First the quantitative evaluation is presented, next we examine some of the patches to get further insights into the nature of the repaired bugs.

\subsection{Quantitative Evaluation}

The calculated metric values of \emph{nDCG} described in Equation~\ref{eq:NDCG} of Section~\ref{sec:method} can be found in Figure~\ref{fig:metrics}. The possible values of the metric ranges from 0.0 to 1.0, a higher metric value means better ranking. We can see that in case of Eslint 217 and Eslint 41 the values are 1.0, this is clearly because the developer fix was ranked to the first place in these cases and irrelevant documents were placed on the end of the similarity list. Based on this metric it is clear that in most cases similarity lists hold their place in ranking patches. Let us discuss further the cases where the metric values are lower.

The \emph{nDCG} metric value reach it's lowest point at the Eslint 94 bug. If we take a look at the subplot (e) at Figure~\ref{fig:sims} we can see that indeed that Doc2vec failed to rank the developer fix at the top of the list. Moreover, most of the patches at the top of the list are incorrect ones, meaning that they hold low relevance. The case of Eslint 100 is different from the previous one. Although it is true that the \emph{nDCG} value of Eslint 94 is 0.67, while for Eslint 100 it is 0.84, compared to others it still seems to be quite low. If we take a look at the rankings at Figure~\ref{fig:sims} we can see that in this case the developer fix is placed on the second place of the ranked list. So the question arises, what causes the low metric value? The answer is quite obvious: the patch which is placed ahead is an incorrect one, decreasing the metric value drastically. The case of Eslint 221 is also interesting: although the developer patch is placed closer to the end of the list than anywhere else, the \emph{nDCG} metric value is not that low. This is due to the fact that in case of this bug the majority of generated plausible patches are semantically the same as the developer fix, resulting in overall higher relevance scores.

Let us now examine the similarity lists in Figure~\ref{fig:sims}. It is cleary visible that in most cases the developer fix has been placed on a prime location in the similarity list. The developer fix is at the top of the list in 3 cases and takes the second place in 4 cases. We would like to note that the ranking of the lists are quite instable for two reasons: (1) the numerical difference is not outstanding between each similarity value, and (2) Doc2Vec fails to give back identical similarity value even though the same documents are compared. Because of these previously mentioned limitations different Doc2Vec model trainings can even result in completely distinct lists of similarities.

\subsection{Qualitative Evaluation}

Due to space constraints we cannot analyze manually every patch, however, we will present two of them in this subsection. We would like to emphasize that our goal is not to use the similarity list as a classifier — which decides whether a patch is correct or not — but to analyze the patches. For each of the bugs described below, the developer fix is listed in the Table~\ref{tab:bugsjs_repairs}. The first patch we are going to examine in more details is the one generated for Eslint 1.

\begin{mdframed}[roundcorner=5pt, backgroundcolor=black!5, innerleftmargin=10pt, innerbottommargin=0pt, innertopmargin=0pt]
	\begin{lstlisting}[language=JavaScript, escapechar=@]
	const name = node.callee.name;            
                            
@\colorbox{patchred}{	- if (name === "Math" || name === "JSON") \{ \hspace{14pt} }@
@\colorbox{patchgreen}{	+ if (name === "Math" | name <= 'S') \{ \hspace{43pt}}@
		context.report(node, "'{{name}}' is not a function.", { name });            
	}
	\end{lstlisting}
\end{mdframed}
\vspace{-3pt}
\mycaption{Listing 1: Original code of Eslint 1 (-), and the most similar automatically generated patch to it (+)}
\vspace{-3pt}

In Listing 1 one can examine the original line with red background and the line that was generated by an APR tool with green background. In this case the developer fix adds another logical testing in the if condition, allowing the \sourcecode{name} variable to have the value \sourcecode{"Reflect"} as well. The modifications which the APR tool made bears no resemblance to this. At first sight it does show greater similarity with the original line than the developer fix, however the generated code is clearly not the best. First let us understand the generated line of code. The \sourcecode{name} variable contains a string value and it is compared with \sourcecode{\textless=} relation to another string value. In JavaScript if both values are strings, they are compared based on the values of the Unicode code points they contain. Meaning that every string which begins with a letter in front of S in alphabetical order will evaluate to \sourcecode{true} otherwise \sourcecode{false} - according to this if the \sourcecode{name} variable has the value \sourcecode{"Reflect"} or \sourcecode{"JSON"} it will evaluate to \sourcecode{true}. So far not that bad. Surprisingly changing the logical \emph{or} (\sourcecode{||}) operator to the bitwise or operator (\sourcecode{|}) does not have any effect here, since the bitwise operation \sourcecode{false | true} results in \sourcecode{1} which converted to boolean evaluates to \sourcecode{true}. Similarly true for every case of the logical operation.

Overall classifying this patch as an incorrect feels a bit ill-judged. If an experienced software developer examines this code modification, he comes to the conclusion that the fix has something to do with the \sourcecode{name} variable. However it might be true that the generated patch is overfitted, it contains valuable information about the repair i.e. gives a hint to the developer which variable might cause the incorrect behaviour.

\begin{mdframed}[roundcorner=5pt, backgroundcolor=blue!5]
\noindent\textbf{Answer to RQ1:} Based on the observed patches, we would recommend a more sophisticated technique to validate patches than plain source code embeddings, because as we have seen the problem itself is more nuanced and complex than a simple true/false classification.
\end{mdframed}

One can argue that this is due to the fact that fixes are often limited to a single line, and in some cases only a single character is affected (eg. \texttt{\textgreater} instead of \texttt{\textless} in an \emph{if} structure). We definitely have to mention that the patches were generated with the use of a single APR tool, it is hard to justify if the conclusions are valid for other tools and multi-line fixes as well. However, defining a threshold and based on this deciding on the correctness of a patch, seems to oversimplify the decision criterion too much. On the other hand, the strive for understandable and simple patches is a reasonable and important aspect of automatic software repair. Generating unreadable patches does not help much with a real-life problem. But if a patch is not too similar to the original program, does it exclusively mean that it is unreadable? On Listing 2 we can observe another code snippet from the Eslint project, but this time we picked the least similar generated patch from the bug Eslint 321. At first glance the two lines seem to be very similar even though in the similarity list it was the last one. The latter does not necessarily mean a big difference, especially if there are very few candidates: in this special case even the last plausible patch shows great similarities with the original program.

\begin{mdframed}[roundcorner=5pt, backgroundcolor=black!5, innerleftmargin=10pt, innerbottommargin=0pt, innertopmargin=0pt]
	\begin{lstlisting}[language=JavaScript, escapechar=@]
 fix: fixer=>fixer.insertTextBefore(node, "\n")
	});           
                            
@\colorbox{patchred}{\scriptsize- else if(tokenBefore.loc.end.line !== node.loc.end.line}@
@\colorbox{patchred}{	\hspace{70pt} \scriptsize\&\& option === "beside") \{ \hspace{42pt} }@
@\colorbox{patchgreen}{\scriptsize+ else if(tokenBefore.loc.end.line - node.loc.start.line}@
@\colorbox{patchgreen}{	\hspace{70pt} \scriptsize\&\& option === 'beside') \{ \hspace{42pt} }@            
    context.report({            
        node,
	\end{lstlisting}
\end{mdframed}
\vspace{-3pt}
\mycaption{Listing 2: Original code of Eslint 321 (-), and the least similar automatically generated patch to it (+)}
\vspace{-3pt}

In case of Eslint 321 the developer fix only changed the \sourcecode{end} word to \sourcecode{start}. This is obviously a small bug and is probably due to developer inattention. Though the automatically generated fix also changed this class member it made further changes. First it changed the double quotes (\sourcecode{"}) to single ones (\sourcecode{'}). Next deleted the strict not equal operator (\sourcecode{!==}) to subtraction (\sourcecode{-}). The first change obviously did not affect the meaning of the \sourcecode{if} structure, but neither did the latter, because if the observed two values are equal and we subtract them, the result is 0, which is evaluated to \sourcecode{false} in JavaScript.

\begin{mdframed}[roundcorner=5pt, backgroundcolor=blue!5]
\noindent\textbf{Answer to RQ2:} The last item in the similarity list can also be a semantically correct one, even though it is less similar to the original program. From this, we can conclude that while similarity-based methods may be suitable for filtering out too many patches, one should use them for classification cautiously bearing in mind the possible misclassification.
\end{mdframed}

In total we examined 465 automatically generated patches as can be seen in Table~\ref{tab:bugsjs_repairs}. From these 13 were syntactically equivalent to the developer fix and 211 semantically. We found that most of the semantic-matched patches were more similar to the original code than others. This behavior can be observed on Figure~\ref{fig:metrics}, where the metric values were calculated using data annotation based on the correctness of each patch. The similarity of the developer fix also tends to be close to the original program as one can see on Figure~\ref{fig:sims}. Our experiments targeted one-line modification and the evaluation was conducted on only one project. These might seem to be limitations, however at the time of writing the paper, there was no available APR tool, which could generate multi-line patches for JavaScript programs. We encourage the interested reader, who is interested in more analysis of the generated patches read the paper of GenProgJS, linked above.

\section{Related Work}\label{sec:related}

To generate repair patches as simple as possible, has already mentioned in many works~\cite{Mechtaev2015,White2017,Tufano2019}. According to a recent study ~\cite{Wang2019} 25.4\% (45/177) of the correct patches generated by APR techniques are syntactically different from developer provided ones. Other approaches also exists, which generate patches by learning human-written program codes ~\cite{Le2016,Kim2013a}. While such approaches have shown promising results, they have recently been the subject of several criticisms ~\cite{Monperrus2014}. 

In automated program repair the evaluation of existing approaches is crucial. Evaluating APR tools based on plausible patches are not accurate, due to the fact of the overfitting issue in test suite-based automatic patch generation. Finding the correct patches among the plausible patches requires additional developer workforce. Liu et al.~\cite{Liu2021} proposes eight evaluation metrics for fairly assessing the performance of APR tools beside providing a critical review on the existing evaluation of patch generation systems.

In a recent study~\cite{Wang2020} benchmarks the state of art patch correctness techniques based on the largest patch benchmark so far and gathers the advantages and disadvantages of existing approaches beside pointing out a potential direction by integrating static features with existing methods. Another work where embedding methods were used for ranking candidates is~\cite{Tian2020}, where beside Doc2Vec and Bert, code2vec and CC2Vec were also applied. In this work they investigate the discriminative power of features. They claim that Logistic Regression with BERT embedding scored 0.72\% F-Measure and 0.8\% AUC on labeled deduplicated dataset of 1,000 patches.

Recommendation systems are not new to software engineering ~\cite{Kochhar2016,Robillard2010,Robillard:2014:RSS}, presenting a prioritized list of most likely solutions seems to be a more resilient approach even in traceability research ~\cite{Kicsi:2018:RAISE:Traceability,Csuvik2019::ISSQ,Csuvik2019::SST}. In a recent work~\cite{Csuvik2020::IBF} candidate patches were ranked based on their similarity and evaluated as a recommendation system. They have proposed a small-scale study on the use of embeddings: leveraging pre-trained natural language sentence embedding models, they claim to have
been able to filter out 45\% incorrect patches from the QuixBugs dataset.

In a recent study~\cite{Le2019} authors assessed reliability of automated annotations on patch correctness assessment. They constructed a gold set of correctness labels for 189 patches through a user study and then compared these labels with automated generated annotations to assess reliability. They found that independent test suite alone might not serve as an effective APR oracle, it can be used to augment author annotation. In the paper of Xiong et al.~\cite{Xiong2018} the core idea is to exploit the behavior similarity of test case executions. The passing tests on original and patched programs are likely to behave similarly while the failing tests on original and patched programs are likely to behave differently. Based on these observations, they generate new test inputs to enhance the test suites and use their behavior similarity to determine patch correctness. With this approach they successfully prevented 56.3\% of the incorrect patches to be generated.

Syntactic or semantic metrics like Cosine similarity and Output coverage~\cite{Le2017} can also be applied to measure similarity, like in the tool named Qlose~\cite{DAntoni2016}. These metrics have several limitations, like maximal lines of code to handle or that they need manual adjustment. On the other hand, the use of document embeddings offers a flexible alternative. Opad~\cite{Yang2017} (Overfitted Patch Detection) is another tool, which aims to filter out incorrect patches. Opad uses fuzz testing to generate new test cases and employs two test oracles to enhance validity checking of automatically generated patches. Anti-pattern based correction check is also a viable approach ~\cite{Tan2016}.

\section{Conclusions}
\label{sec:conclusion}

Patch validation in the APR domain is a less explored area, which holds great potential. Filtering out incorrect patches from the set of plausible programs is an important step forward to boost the confidence towards APR tools. In this paper we experimented with a similarity-based patch filtering approach and conducted a quantitative and qualitative evaluation of it. The similarity between patches was calculated with the use of source code embeddings produced by Doc2Vec. Although the applied approach may be useful when a high number of plausible patches are present, we found that plain source code embeddings fail to capture nuanced code semantics, thus a more sophisticated technique is needed to correctly validate patches. We expect that a more complex language understanding model may be advantageous in deciding whether a patch is correct or not. In future work, we wish to explore whether the use of deep learning techniques constitutes additional advantages.

\vfill\null

\section*{Acknowledgement}
he research presented in this paper was supported in part by the ÚNKP-20-3-SZTE and ÚNKP-20-5-SZTE New National Excellence Programs, by grant NKFIH-1279-2/2020 and by the Artificial Intelligence National Laboratory Programme of the Ministry of Innovation and the National Research, Development and Innovation Office. László Vidács was also funded by the János Bolyai Scholarship of the Hungarian Academy of Sciences.

\bibliographystyle{IEEEtran}

\end{document}